\documentclass[conference]{IEEEtran}

\IEEEoverridecommandlockouts

\usepackage{amsmath}
\usepackage{url}
\usepackage{epstopdf}
\usepackage{color}
\usepackage{hyperref}
\usepackage{amsopn}
\usepackage{amssymb}
\usepackage{subfigure}
\usepackage{eurosym}

\usepackage{xcolor}
\usepackage[pdftex]{graphicx}
\usepackage{wrapfig}

\DeclareGraphicsExtensions{.pdf,.png}

\begin{document}

\title{Cyber Resilience: principles and practices%
\thanks{%
\protect\begin{wrapfigure}[3]{l}{.9cm}%
\protect\raisebox{-12.5pt}[0pt][0pt]{\protect\includegraphics[height=.8cm]{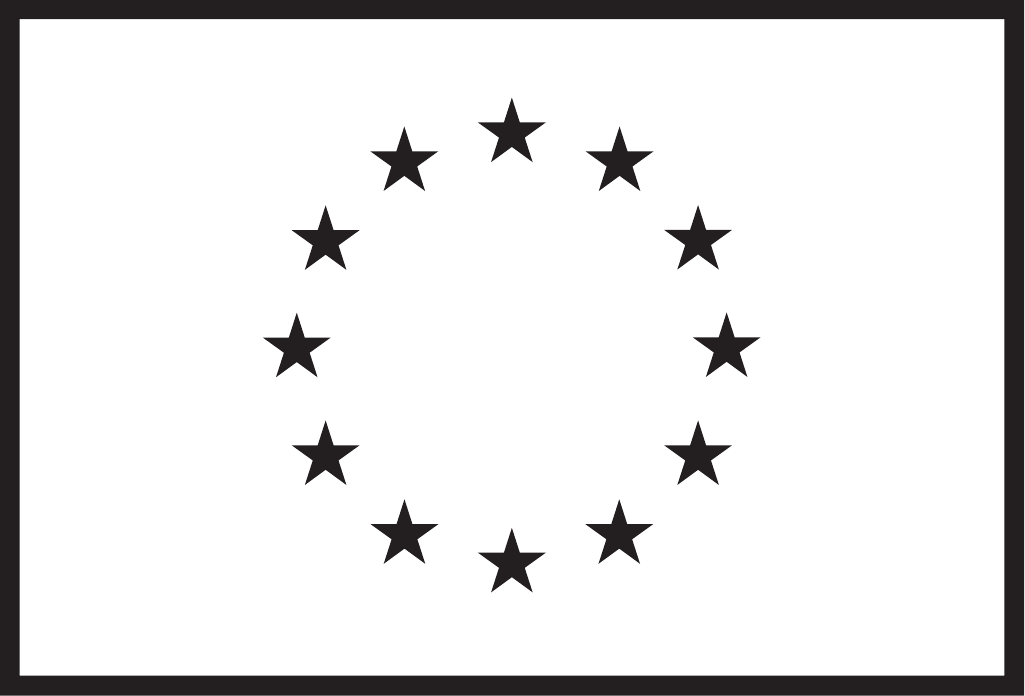}}%
\protect\end{wrapfigure}%
This project has received funding from the European Union's Horizon 2020 research and innovation programme under grant agreement no. 786698. The work reflects only the authors' view and the Agency is not responsible for any use that may be made of the information it contains.}}

\author{\IEEEauthorblockN{%
Emanuele Bellini\IEEEauthorrefmark{1},
Franco Bagnoli\IEEEauthorrefmark{2},
Alexander A. Ganin\IEEEauthorrefmark{4},
Igor Linkov\IEEEauthorrefmark{5}\\
\vspace*{4pt}}

\IEEEauthorblockA{\IEEEauthorrefmark{1}Mathema s.r.l., Italy; Khalifa University Center of Cyber-Physical System, UAE.  Email: emanuele.bellini@ieee.com}
\IEEEauthorblockA{\IEEEauthorrefmark{2}Department  of Physics and Astronomy and CSDC,  University of Florence. Email: franco.bagnoli@unifi.it}
\IEEEauthorblockA{\IEEEauthorrefmark{4}University of Virginia, USA. Email: aag2k@virginia.edu}
\IEEEauthorblockA{\IEEEauthorrefmark{5}Engineer Research and Development Center, U.S. Army Corps of Engineers, USA. Email: Igor.Linkov@usace.army.mil}

}

\maketitle

\begin{abstract} 
Cyber Resilience is complex cyber physical systems is becoming even more important in fully connected world. This work aims at formalizing the concept and exploring current promising practices for resilience quantification and assessment in the context of Internet of Everything. 
In particular IoE has been considered a network of heterogeneous devices are exposed to cyber attack. Cyber Resilience paradigm provides a comprehensive framework to define evidence driven strategies to balance between the need of reducing the malware outbreak while maintaining the network functionalities at an acceptable level. In this war game a number of assets and actors come into play with different rules and contributions to the overall resilience. In this respect a resilience ontology able to reconciliate different perspective and definitions in resilience domain is provided. 
\end{abstract}


\maketitle

\section{Introduction}
Our society is becoming even more connected. 

The environment is increasingly pervaded by devices capable of interacting with the environment itself, other devices and human beings.

The IoE can be considered a
natural development of the IoT concept. In fact, while ”Things” are related to connect
physical-first objects, IoE extends this view comprising the following four key elements
including all sorts of possible connections:
a) People: Considered as end-nodes connected across the Internet to share knowledge,
information, opinions, decisions, behaviors, and activities.
b) Things: Physical sensors, devices, actuators and other items generating data or receiving
information from other sources.
c) Data: Raw data analyzed and processed into useful information to enable intelligent
decisions and control mechanisms (e.g., Human behaviors on the ground).
d) Processes: Leveraging connectivity among data, things, and people to add value.
Thus IoE establishes an end-to-end ecosystem of connectivity where people with their
relationships, social collaborations, and grouping dynamics represent an integral part.

Our environment is becoming saturated with computing, sensing and communication devices that interact among themselves, as well as with humans: virtually everything is enabled to generate data and respond to appropriate stimuli (Internet of Everything). 

According to ERICSSON Mobility report \cite{ericson}, the number of connected IoT devices is expected to exceed the number of mobile phones, and so is anticipated to revolutionize the way we do business, communicate, and live. This Cyber Physical World (CPW) convergence results in humans being deeply immersed in the information flows from the physical to the cyber world, and vice-versa \cite{conti}. The services being offered via platforms for enabling the IoT vision are becoming highly pervasive, ubiquitous, and distributed; machines, objects, and services become more intelligent and create a large-scale decentralized pool of resources interconnected by highly dynamic networks.
Such technological evolution is also making our society vulnerable to new forms of threats and attacks exploiting the complexity and heterogeneity of IoT networks, therefore rendering the cyber-security amongst the most important aspects of a networked world. Indeed, the security and management of the vast volumes of data generated, transmitted, and stored by smart environments and devices is still difficult to achieve. According to \cite{delic}, at the conceptual level, IoT can be considered as a hierarchical tree of self-regulating and self-managing sub-systems and devices with high-density connectivity and very low system latencies. While this is desirable when the system is in a stable regime, easy and fast propagation of perturbances may create important stability problems. Considering possible crashes and cyber-attacks, and assuming the under-specified nature of performance conditions in the system \cite{ryan}, a certain level of epistemic and aleatory uncertainty must be taken into account as a contribution to critical events \cite{bellini}. In this respect, a resilience approach to IoT seems to be the best option thanks to its capacity to overcome the weakness of the current risk and efficiency-based approaches in complex socio-technical systems safety and security management in addressing the so called “unknown unknowns” \cite{park}. This work extends the current research on cyber security performed in the context of Cyber Trust EU project where IoT vulnerabilities and attacks are studied to improve the response capability of a cyber physical system\cite{crawling}, \cite{ifip}.

We review the challenges associated with IoT in Section 2 and detail the cyber resilience formalization in IoT in Section 3. Next, we  formalize the  problem and the approach in section 4 while the introduction of risk perception with memory model is formalized and simulated in section 5; Section 6 includes conclusion and next steps. 

\section{Background}	

As networked devices become ubiquitous, cyber-attacks will become more frequent and even more sophisticated. There are already numerous recent examples of cyber-attacks that exploit the Internet-connected appliances, such as refrigerators, televisions, cameras, and cars, in order to e.g. perform denial-of-service (DoS) or distributed DoS (DDoS) attacks of unprecedented scales, spy on people in their office/homes, and take over (hijack) communication links thus delivering full control of anything that is remotely controlled, like drones and vehicles, to cyber-criminals. For instance, in the health sector, potentially deadly vulnerabilities have been found in a large number of medical devices, including insulin pumps, CT-scanners, implantable defibrillators, and x-ray systems. Cyber-security incidents targeting critical information infrastructures (CIIs), which provide the vital functions that our society depends upon, are expected to have a significant negative economic and societal impact in the next decade and should be considered global risks. A review report recently published by ENISA estimates that the average annual losses due to cyber-crime among the European Union (EU) countries is 0.41\% of their gross domestic product (GDP); in some countries (Germany and Netherlands) the losses exceed 1.50\% of the GDP leading to annual costs in the range \euro 425K -- \euro 20M per company. Amongst the CII sectors within the EU, significantly affected ones are the financial, information and communication technology (ICT), health, transport, energy, and public. DoS/DDoS, targeted, and web-based attacks, along with ransomware, have been reported to be amongst the most common types of cyber-attacks \cite{symantec}. The availability of botnets-for-hire led to a noticeable increase in DDoS attacks, and it is very likely to see the IoT to further facilitate the formation of such botnet armies. The number of zero-day, i.e. previously unknown and immediately exploitable, vulnerabilities discovered, has roughly doubled in the last few years; given their value, a rather mature black market has evolved that allows these vulnerabilities to be employed (until their exposure) in sophisticated targeted attacks. Hence, the deployment of proactive security and threat intelligence gathering/sharing systems could prove to be efficient in preventing such types of cyber-attacks. Web based attacks exploit websites’ vulnerabilities to infect users or gain access to sensitive private data, due to misconfigurations, usage of no/weak security protocols, or lack of proper patch management procedures. More than 75\% of the websites have unpatched (critical) vulnerabilities, which may still be exploited several months after the vulnerability is revealed. This allows cyber-criminals to put little or even trivial effort in taking over control of the systems that can afterwards be utilized in numerous ways \cite{krebs}.
In the case of IoT, due to the security problems arising from embedded devices and other legacy hardware, whose flawed design (such as, the use of hardcoded administrative passwords) or their poor configuration allows cyber-criminals to easily compromise them in order to form powerful botnets and launch DDoS attacks. Most importantly, there is often no efficient way to patch those devices. Many such IoT devices can be located by using new search engines, for example, SHODAN (www.shodan.io), and this gives cyber-criminals the opportunity to exploit any existing vulnerabilities on a large scale. Important questions to consider in this area include: a) How to prevent large-scale vulnerabilities in IoT devices; b) How to prevent existing vulnerabilities in IoT devices from being exploited on a large scale; c) How to stop a large-scale propagation of the attacks after a vulnerability has been exploited while maintaining the network functionality at an acceptable level of performance (degraded mode). 
Therefore, there has been an increasing interest in shifting emphasis from configuring the compliance of devices with respect to security practices towards measuring compliance. If such information could be coupled with traffic analysis techniques, then the above approach would allow computing an accurate threat (or trust) score of the devices accessing a corporate network, and could counter threat actors and their methods by providing greater assurance that a device is functioning properly and uses adequate security mechanisms.
Compromised IoT devices may exhibit arbitrary behavior, and hence communications from any such device should be quarantined, or even rejected, by intrusion prevention systems. Security services such as deep packet inspection, protocol analysis, and data analytics can be implemented in order to identify anomalies in the data generated by IoT devices and other potential cyber-threats. In the current networking paradigm of IoT ecosystems, where lightweight endpoint devices rely on some central cloud server in order to be authenticated and identified, network operators can provide IoT service providers the aforementioned security services. Due to the vast volumes of traffic that IoT devices are expected to generate in the near future and the employment of peer-to-peer (P2P) communications, the above security services should be distributed across all the devices with adequate processing capabilities.

\section{Cyber Resilience in IoT}
In the domain of IoT, the concept of resilience is very new and its understanding varies according to the perspective adopted about the nature of IoT. For instance, in \cite{sherratt}, a resilient IoT system should deal with a number of threats occurring when a system is deployed on the Internet \cite{sherratt2} in order to safely, and quickly recover to provide normal service. Literature on self-healing systems goes even further, not only recovering normal behavior, but also addressing the vulnerability that led to faulty behavior \cite{cardinal}. Thus, resilience relates to fault-tolerant systems, dependable and trustable systems, and reliable and available systems, where each category of applications adopts its specific terminology and angle of interest.  
A common approach for resilience evaluation and assessment is required for IoT literature. IoT can be considered as techno-social systems, combining infrastructures, devices, and people \cite{delic}. Thus, to address IoT resilience, it is necessary to study the resilience of complex systems (networks). Resilience was defined as a property of networked system in \cite{ganin}. This framework adopts the definition of resilience given by the US National Academy of Sciences \cite{nrc} and further discussed in  \cite{nature}: 
\begin{itemize}

\item •	Plan/Prepare: Lay the foundation to keep services available and assets functioning during a disruptive event (malfunction or attack).
\item •	Absorb: Maintain most critical asset function and service availability while repelling or isolating the disruption.
\item •	Recover: Restore all asset functions and service availability to their pre-event functionality.
\item •	Adapt: Using knowledge from the event, alter protocol, configuration of the system, personnel training, or other aspects to become more resilient.
\end {itemize}

The relation between system resources, the capacities that they provide, and variability can only become meaningful when placed in the context of the interdependencies that are on the one hand used to enhance resources and capacities but also, on the other hand, require an allocation of resources to be maintained \cite{esrel},\cite{esrel1}. Interdependencies are the means through which system functions can act in order to secure the envisaged levels of resources needed to fulfill their purposes. To this end, the principles of network science are proposed as a tool to quantify resilience by supporting system interdependencies in view of the volume and nature of available resources and the capacities that these provide. At the operational side, we assume that resilience can be quantified as a property of an interdependent network system. As such, resilience assessment would identify critical functionality of the system \cite{bellini,bellini2,trento} and evaluate the temporal profile of system recovery in response to adverse events, while resilience management would allow comparative evaluation of cross-domain management alternatives. Resilience assessment may be approached with network theoretic frameworks \cite{ganin}. A network is composed of a set of nodes $(N)$, connected by a set of links $(L)$. The specification of $N$ and $L$ includes characteristics relevant to resilience (e.g., capacity, geographical location, weight, temporal response) of each node. The adverse events and failures can percolate over a network in different ways. The network control $C$ captures the system's temporal evolution, including adaptive algorithms, and can also be defined as a way to manage temporal changes in the network. Ultimately, the system must maintain its critical functionality $(CF)$ that reflects the state of the system as a characteristic (or a weighted sum of such characteristics) that changes with time and is of interest to the user of the model. In \cite{ganin}, Resilience $(R)$ is then defined as a composite function of nodes, links and control with respect to the critical functionality of the system and a class of disturbances as
\begin{equation}
R=R_{CF,P}=f(N,L,C)
\end{equation}
Due to the very complex nature of networked systems and the large number of variables defining their states, it is not possible to obtain a closed-form expression for R. Thus, to obtain a quantification of resilience, several approaches and strategies are  adopted depending on the level of rigor needed and resources available \cite{tiered} such as \cite{bellini,bellini2,nato,ganin_resilience_2019,gisladottir_resilience_2017,massaro_resilience_2018, esrel}.
In particular, in IoT, the best approach is represented by simulations \cite{ganin}, each of which represents a certain network percolation \cite{percolation} scenario from an infinite set of possible scenarios of the network’s evolution. For each simulation, it is possible to calculate the average value of the critical functionality at every time step. To follow this experimental approach, we also introduce the control time $T_C$ as the number of time steps for which the value of CF is to be determined. Essentially, the control time represents a time period set by stakeholders during which we estimate the resilience of the system. Resilience is then quantified as the integral of the system functionality during a disruption, normalized to its normal functionality. In the present article we adopted the same approach, using simulation to assess the resilience of the IoT network using a risk perception based approach in epidemic spreading model in a given time window. 

%
%

\begin{figure}[t!]
\centering
{\includegraphics[width=0.6\columnwidth]{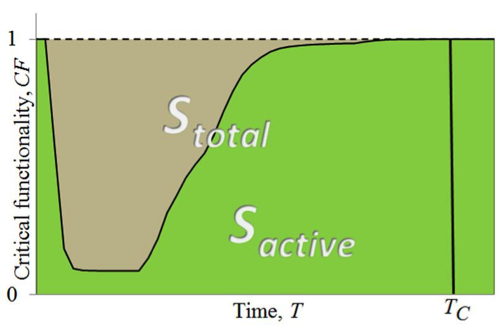}}

\vspace{-1em}
\caption{Typical profile of the critical functionality: in a networked system where the nodes are perturbed and becoming inactive, the resilience R can be quantified as the integral of $CF: S_{active}/S_{total}$}
\vspace{-1.5em}
\end{figure}

\section{Approach}
Similarly to epidemic spreading processes, in the mobile IoT environment, malware may be transmitted through opportunistic networks based on devices proximity \cite{mcom,wimeh}. For example, mobile phone viruses may spread through multimedia messages or Bluetooth \cite{wang}. Such proximity-based ad-hoc networks closely resemble people interaction within a population. Analogous to travel restrictions for epidemic containment, actions to prevent the spread of malware in the IoT include device lockdowns, restrictive security settings, as well as social domain countermeasures, such as corporate policies limiting the use of equipment. These countermeasures, derived from a risk management approach, are due to the fact that the resources to detect an attack on every single device in the IoT to react selectively are often not available. Thus, in order to avoid attack propagation, entire sub-networks are disconnected from the infrastructure generating costs for service outage. This leads to the classical problem of balancing the requirements for users’ convenience and performance with the devices’ security. The emerging field of resilience quantification could provide means for identifying an appropriate response to a malware outbreak given the scale of its adverse impact. Resilience management expands the purview of individual risks by monitoring the response of an entire system to disturbances and their aftermath, including mitigation efforts \cite{sheffi}. In particular the risk of infection/attack needs to be balanced in order to reduce the malware outbreak while maintaining the network functionality at an acceptable level.
According to the theoretical epidemiology field,  there is a strong relationship between the infection probability $\tau$,  the average number of contacts $\langle k \rangle$ and its variance $\langle k ^2\rangle$: the critical value $\tau_c$ for the onset of an epidemic is 
\[
\tau_c = \frac{\langle k \rangle }{\langle k^2 \rangle}\simeq \langle k \rangle ^{-1}
\]
for sharp-distributed networks~\cite{risknet}. The wide distribution of IoT devices as well as human contacts can generally be approximated with a scale-free distribution with diverging variance, thus it is infeasible to control epidemics only through the reduction of infection probability. 
Moreover, security assessments typically scan for known issues while there is a chance that a dedicated attacker finds a previously unknown issue, a so called zero-day \cite{yin}.  There are several metrics in security from which it is possible to derive a probability of an attack such as metrics of system vulnerabilities, metrics of defense power, metrics of situations, and metrics of attack or threat severity \cite{survey,ganin_multicriteria_2017}. However, the metrics mentioned above are not ideal when faced with unknown threats. To this end, as introduced in previous works~\cite{riskperception}, \cite{complex} and \cite{compeng}, risk perception in epidemic spreading applied in IoT networks can be used to support countermeasures in presence of unknown risks while trading off between the costs  (functionality reduction) generated by the damage and the costs generated by the application of countermeasures. The main assumption is that the knowledge about the diffusion of the disease among neighbours (without knowing who is actually infected) effectively lowers the probability of transmission (the effective infectiousness). Moreover in a highly connected network such as the IoT, the percolation/infection threshold should not be presented and the attack can be propagated to the entire network without the possibility of mitigate respond maintaining a certain level of functionality (e.g. graceful degradation). However in \cite{riskperception}, \cite{complex} and \cite{compeng} it has been shown that, if the  risk perception at the node level (local) is increased, it is possible to stop the attack propagation even in a scale-free network. 

In a real case this perception of risk is "triggered" by the "global" percentage of the infected. Thus, if the probability of infection depends on the global percentage of infected, the article aims at showing that with a memory mechanism it is possible to obtain a resilient behavior in a IoT network minimizing the loss of functionality during an attack.
In the analysis, we considered the possibility that the information coming from the cyber-physical contacts is also influenced by the (mobile) social contact networks~\cite{virtual_inf, virtual_inf1, virtual_inf2}. In the case of the IoT, it is vital to quickly assess the epidemic threshold for a given network (that may change in time), with real-time estimates of the infection probability, that may change from node to node. We present here a self-organized method (first introduced in Ref~\cite{bagnoli_rech} and extended in \cite{compeng}, \cite{complex} ) that can be applied in such situations.

\section{Risk perception with Memory}

As defined in \cite{compeng} and \cite{complex},  the problem can be formalised as follows. There is a set of $N$ nodes $x_i$ in IoT network, that can stay in two states: 0 for ``healthy'' and 1 for ``tainted'' (contaminated, attacked). The node $i$ processes information coming from other nodes $j$, defined by an adjacency matrix $a_{ij} = 1 (0)$ for connected (disconnected) nodes. This information is then propagated to other $j$ nodes, again defined by $a_{ji}$.  A node $i$ in tainted status can ``infect'' other nodes with a probability $\tau$, that may depends on several known and unknown factors. The nodes can also respond to the attack themselves by checking the information they own with a central server that maintains the world state of the system. However, this operation takes time and bandwidth, and it has therefore a certain cost. For instance a node can suspend its operation while waiting for a validation check reducing the level of network functionality and thus its resilience. We define the input connectivity of node $i$ as $k_i = \sum_j a_{ij}$. The use of the concept of risk perception in IoT networks  as a method to decide if it is preferable to suspend the information processing reducing the functionality (and the logical connectivity) of the network or to take the risk of elaborating false data or allowing attack propagation has been introduced in \cite{compeng}, \cite{complex}. The nodes in IoT, randomly, could check the correctness of incoming information against a central server that keep the status of the entire network. However, this procedure has a cost in terms of both time and bandwidth that during a cyber attack cannot be acceptable. So, the lower is the infection probability, the higher the cost. 
We have shown that, increasing the level of local risk perception, it is possible to stop the epidemics also in  a scale-free network \cite{riskperception}, \cite{complex} and \cite{compeng}. However, this effect comes at the price of large cost, so we developed techniques for automatically detecting the infection threshold, so to apply the minimum effort needed to stop the epidemics. We want to show here that it is possible to have a system able to automatically trigger the needed level of alarm in order to stop the propagation of an attack (epidemics).
    
      \begin{figure}[t]
        \centering
        \begin{tabular}{cc}
            (a) & (b)\\
            \includegraphics[width=0.48\linewidth]{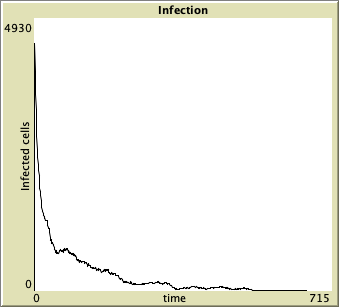} &
            \includegraphics[width=0.48\linewidth]{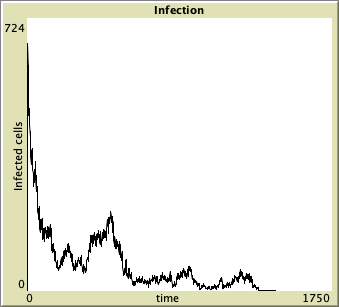}\\
        \end{tabular}
        \vspace{-1em}
        \caption{Plot of the number of infected cells in a $100\times 100$ lattice, with initial infection level equal to $0.5$. (a) $J=0$, $\tau\simeq\tau_c=0.135$. (b) $\tau=0.5$, $J = 1.05$. }
        \label{fig:tau-J}
        \vspace{-1.5em}
    \end{figure}
    We use a SIS (susceptible-infected-susceptible) model of infection \cite{iobt}, on a square lattice. The state of each cell $s_{ij}$ can be either 0 (healthy) or 1 (infected). Infected sites recovery in one step, but may infect neighbouring sites with probability $\tau$. The infection can come independently from any infected neighbours (site percolation), so that the probability $p(n)$ of being infected if one is surrounded by $n$ infected neighbours is  
    \[
    p(n) = 1- (1-\tau)^n. 
    \]
We consider the case of nearest and next-to-nearest connections, so that a cell is connected to 8 neighbours. In this case the critical infection probability is $\tau_c \simeq 0.136$. 
    
    The local risk perception is implemented by  replacing the bare infection probability $\tau$ with 
    \[
        \tau \exp(-J  n). 
    \]
    By increasing $J$, one can stop the infection even for $\tau > \tau_c$. For instance, setting $\tau=0.5$, it is still possible to stop the epidemics by setting $J=1.05$, as shown in Fig.~\ref{fig:tau-J}. In \cite{riskperception}, \cite{complex} and \cite{compeng} we illustrated how these epidemic thresholds can be obtained in a self-consistent way. This procedure can indeed lower the cost of recovering from the infection: when there is no infection, $n=0$ and therefore the cost, which is proportional to $1-\tau$, is lower than $1-\tau_c$.  However, we face a relevant problem: the number of infected neighbours is actually not observable\dots unless one checks them against the central server, which is exactly what one wants to avoid. We can however use a global quantity: since the central server has the measure of the level of the infection for the entire network, from the number of tainted messages received by the checking mechanism, it can issue an adequate alarm level. 
    
     \begin{figure}[t]
       \centering
       \begin{tabular}{cc}
           (a) & (b)\\
        \includegraphics[width=0.48\linewidth]{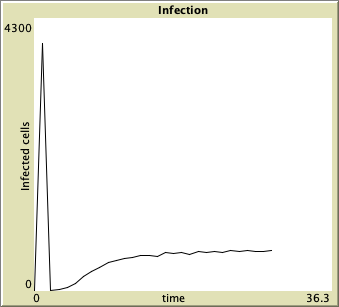} & 
        \includegraphics[width=0.44\linewidth]{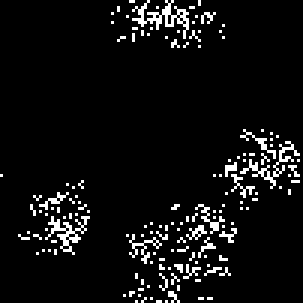} 
       \end{tabular}
       \vspace{-1em}
        \caption{Here $J=0$, $W=0.002$ in a lattice of $100\times 100 cells$. we started with a fraction of 0.1 randomly infected cells. (a) the time plot of the number of infected cells, one can see an immediate drop followed by recovering. (b) The lattice configuration at time 30, the few remaining infected cells after the first clearing gave rise to new outbreaks. }
        \label{fig:w002mem0}
        \vspace{-1.5em}
    \end{figure}
    Let us thus modify the model so that the modulation of the infection probability is 
    \[
    \tau \exp(-J n - W N),
    \]
    where $N$ is the total number of infected cells ($J$ and $W$ can be rescaled with the total neighbourhood and lattice size, if needed). We can set provisionally $J=0$ and examine the role of $W$. We fix $\tau=0.5$. Although the factor $W$ is well able to control a massive outbreak, it appears to be quite ineffective for effectively stopping the epidemics. Specifically, if we start from a certain percentage of infected nodes, their number immediately falls to a small number, but then this factor becomes negligible and the infection recovers, as shown in Fig.~\ref{fig:w002mem0}. 
    In order to investigate this problem, let us  start the simulation with just one infected cell, in order to reproduce an infection episode. In principle, since in the case of a single infected cell, the effective infection probability is $\tau\exp(-W)$, it should be sufficient to equate it to $\tau_c$, getting $W_c = \ln(\tau/\tau_c)\simeq 1.3$ for $\tau=0.5$, but in practice this is not sufficient, since in many cases the number of infected cells grows, and therefore they settle to an asymptotically small, but non-zero, number. In practice, $W$ has to be at least $0.2$ in order to stop the epidemics. 
    
    \begin{figure}[t]
       \centering
\begin{tabular}{cc}
    (a) & (b)\\
        \includegraphics[width=0.48\linewidth]{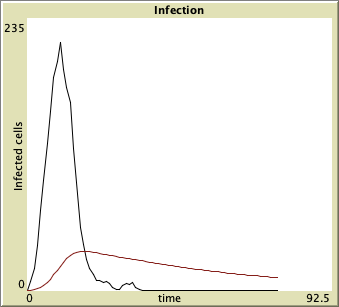} & 
        \includegraphics[width=0.48\linewidth]{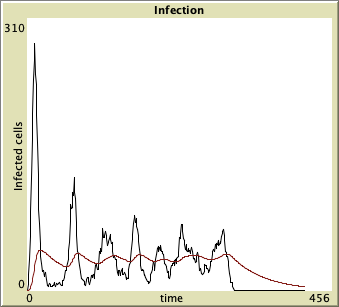} 
        \end{tabular}
        \caption{Time plot of the infection level (black) and risk perception $\Omega$ (red) for $\mu=0.02$. (a) $W=0.05$, (b) $W=0.035$. }
        \label{fig:w-mem}
        \vspace{-1.5em}
    \end{figure}
    
A solution is that of inserting a memory effect. We replace $W$ by a quantity $\Omega$, that evolves in time as
    \[
        \Omega(t+1) = (1-\mu) \Omega(t) +  \mu W(t),
     \]
where $\mu$ is related to the memory characteristic time (if $W=0$ $\Omega$ vanishes as $\Omega(t)=\Omega_0 (1-\mu)^t\simeq\Omega_0 \exp(-\mu t)$).   
     
 As shown in Fig.~\ref{fig:w-mem}, one can see that with the memory effect, the risk perception persists long enough to stop the infection with quite smaller values of the risk perception factor $W$, although the price to pay is the large outbreak at beginning. For small values of $W$ one can see several small outbreaks (Fig.~\ref{fig:w-mem}-b) before eradication. Thus, the level of functionality (cyber resilience) can be considered as the opposite of the number of infected nodes (damage). In Fig.~\ref{fig:w-mem}- is then represented the typical case where, due to the memory effect, an initial large outbreak is automatically recovered. Clearly, the model can be quite improved, looking for the optimal combination of global and local alarm levels and memory options, optimization that depends on the network structure. 

\section{Conclusions}
IoT networks are vulnerable to cyber attacks including denial-of-service (DoS) and jamming attacks. These adversarial behaviors disrupt communications between devices, which leads to link removals in the IoT network. Therefore, to maintain the functionality of IoT networks, they need to be made resilient to malicious attacks.
The results of our simulations allow us to examine and quantify the impact of increased fear as well as communication restrictions on malware outbreaks from risk and resilience perspectives. We quantify risk as the total number of impacted devices during the outbreak. There is a balancing aspect to be considered in risk and resilience evaluation because information to raise fear/awareness has the potential to reduce both the number of infected cases as well as functionality and therefore overall resilience of the IoT network. The severity of communication restrictions required to achieve a real risk reduction is not justified unless the risk of a malware spread is particularly high. While risk management is crucial to epidemic control what is not accounted for, generally, is that the restrictions imposed by risk reduction can be detrimental to the normal IoT functioning and create new risks themselves. Disconnecting the region compromises connectivity; mobility of resources to the affected area is of critical value for the immediate local control of the outbreaks and for preventing its further spread. Cyber resilience as a concept can help us balance risk reduction with the critical functions of a system that allow it to plan for adverse events, absorb stress, recover and predict and prepare for future stressors in order to adapt to their potential threats. Any decision-making process for epidemic containment should take into account complex cost and benefit implications.
We think that the above concepts, can represent a new perspective in IoT cyber security management in the context of increasing complexity and limited resources.

\bibliography{biblio1}{}

\begin{thebibliography}{10}

\bibitem{ericson}
``Ericsson, mobility report: on the pulse of the networked society, interim
  update, sep. 2016.''

\bibitem{conti}
M.~Conti, S.~K. Das, C.~Bisdikian, M.~Kumar, L.~M. Ni, A.~Passarella,
  G.~Roussos, G.~Tröster, G.~Tsudik, and F.~Zambonelli, ``Looking ahead in
  pervasive computing: {Challenges} and opportunities in the era of
  cyber–physical convergence,'' {\em Pervasive and Mobile Computing}, vol.~8,
  no.~1, pp.~2--21, 2012.

\bibitem{delic}
K.~A. Delic, ``On resilience of {IoT} systems: The internet of things (ubiquity
  symposium),'' {\em Ubiquity}, vol.~2016, no.~February, pp.~1:1--1:7, 2016.

\bibitem{ryan}
J.~R. Wilson, B.~Ryan, A.~Schock, P.~Ferreira, S.~Smith, and J.~Pitsopoulos,
  ``Understanding safety and production risks in rail engineering planning and
  protection,'' {\em Ergonomics}, vol.~52, no.~7, pp.~774--790, 2009.

\bibitem{bellini}
E.~Bellini, P.~Ceravolo, and P.~Nesi, ``Quantify {Resilience} {Enhancement} of
  {UTS} through {Exploiting} {Connected} {Community} and {Internet} of
  {Everything} {Emerging} {Technologies},'' {\em ACM Transactions on Internet
  Technology}, vol.~18, no.~1, pp.~1--34, 2017.

\bibitem{park}
J.~Park, T.~P. Seager, P.~S.~C. Rao, M.~Convertino, and I.~Linkov,
  ``Integrating {Risk} and {Resilience} {Approaches} to {Catastrophe}
  {Management} in {Engineering} {Systems}: {Perspective},'' {\em Risk
  Analysis}, vol.~33, no.~3, pp.~356--367, 2013.

\bibitem{crawling}
S.~Shiaeles, N.~Kolokotronis, and E.~Bellini, ``Iot vulnerability data crawling
  and analysis,'' in {\em 1st IEEE Service workshop on Cyber Security and
  Resilience in IoT}, 2019.

\bibitem{ifip}
C.~Constantinides, S.~Shiaeles, B.~Ghita, and N.~Kolokotronis, ``A novel online
  incremental learning intrusion prevention system,'' in {\em 10th IFIP
  International Conference on New Technologies, Mobility and Security}, 2019.

\bibitem{symantec}
``Symantec, internet security threat report, vol. 21, apr. 2016.''

\bibitem{krebs}
B.~Krebs, ``The scrap value of a hacked {PC}, revisited,'' Oct. 2012.

\bibitem{sherratt}
E.~Sherratt, ``Intelligent {Resilience} in the {IoT},'' in {\em {SDL} 2017:
  {Model}-{Driven} {Engineering} for {Future} {Internet}} (T.~Csöndes,
  G.~Kovács, and G.~Réthy, eds.), pp.~46--60, Springer International
  Publishing, 2017.

\bibitem{sherratt2}
E.~Sherratt, I.~Ober, E.~Gaudin, P.~Fonseca~i Casas, and F.~Kristoffersen,
  ``{SDL} - {The} {IoT} {Language},'' in {\em {SDL} 2015: {Model}-{Driven}
  {Engineering} for {Smart} {Cities}} (J.~Fischer, M.~Scheidgen,
  I.~Schieferdecker, and R.~Reed, eds.), pp.~27--41, Springer International
  Publishing, 2015.

\bibitem{cardinal}
P.~M.~D. Scully, {\em {CARDINAL}-{Vanilla}: immune system inspired
  prioritisation and distribution of security information for industrial
  networks}.
\newblock Ph.{D}. {Thesis}, Aberystwyth University, Aberystwyth, UK, 2016.

\bibitem{ganin}
A.~A. Ganin, E.~Massaro, A.~Gutfraind, N.~Steen, J.~M. Keisler, A.~Kott,
  R.~Mangoubi, and I.~Linkov, ``Operational resilience: concepts, design and
  analysis,'' {\em Scientific Reports}, vol.~6, p.~19540, 2016.

\bibitem{nrc}
{National Research Council}, ``Disaster resilience: A national imperative,''
  {\em The National Academies Press}, 2012.

\bibitem{nature}
I.~Linkov, T.~Bridges, F.~Creutzig, J.~Decker, C.~Fox-Lent, W.~Kröger, J.~H.
  Lambert, A.~Levermann, B.~Montreuil, J.~Nathwani, R.~Nyer, O.~Renn,
  B.~Scharte, A.~Scheffler, M.~Schreurs, and T.~Thiel-Clemen, ``Changing the
  resilience paradigm,'' {\em Nature Climate Change}, vol.~4, no.~6,
  pp.~407--409, 2014.

\bibitem{esrel}
E.~Bellini, P.~Nesi, L.~Coconea, E.~Gaitanidou, P.~Ferreira, A.~Simoes, and
  A.~Candelieri, ``Towards resilience operationalization in urban transport
  system: The resolute project approach,'' in {\em Proceedings of the 26th
  European Safety and Reliability Conference, ESREL 2016}, 2017.

\bibitem{esrel1}
P.~Ferreira and E.~Bellini, ``Managing interdependencies in critical
  infrastructures — a cornerstone for system resilience,'' in {\em
  Proceedings of {European Safety and Reliability Conference} 2018},
  (Trondheim, Norway), Taylor \& Francis, June 2018.

\bibitem{bellini2}
E.~Bellini, L.~Coconea, and P.~Nesi, ``A functional resonance analysis method
  driven resilience quantification for socio-technical systems,'' {\em IEEE
  Systems Journal}, 2019.

\bibitem{trento}
E.~Bellini, P.~Nesi, G.~Pantaleo, and A.~Venturi, ``Functional resonance
  analysis method based-decision support tool for urban transport system
  resilience management,'' in {\em 2016 {IEEE} {International} {Smart} {Cities}
  {Conference} ({ISC}2)}, (Trento, Italy), pp.~1--7, IEEE, 2016.

\bibitem{tiered}
I.~Linkov, C.~Fox-Lent, L.~Read, C.~R. Allen, J.~C. Arnott, E.~Bellini,
  J.~Coaffee, M.-V. Florin, K.~Hatfield, I.~Hyde, W.~Hynes, A.~Jovanovic,
  R.~Kasperson, J.~Katzenberger, P.~W. Keys, J.~H. Lambert, R.~Moss, P.~S.
  Murdoch, J.~Palma-Oliveira, R.~S. Pulwarty, D.~Sands, E.~A. Thomas, M.~R.
  Tye, and D.~Woods, ``Tiered {Approach} to {Resilience} {Assessment}: {Tiered}
  {Approach} to {Resilience} {Assessment},'' {\em Risk Analysis}, vol.~38,
  no.~9, pp.~1772--1780, 2018.

\bibitem{nato}
I.~H{\"a}ring, G.~Sansavini, E.~Bellini, N.~Martyn, T.~Kovalenko, M.~Kitsak,
  G.~Vogelbacher, K.~Ross, U.~Bergerhausen, K.~Barker, and I.~Linkov, ``Towards
  a {Generic} {Resilience} {Management}, {Quantification} and {Development}
  {Process}: {General} {Definitions}, {Requirements}, {Methods}, {Techniques}
  and {Measures}, and {Case} {Studies},'' in {\em Resilience and {Risk}}
  (I.~Linkov and J.~M. Palma-Oliveira, eds.), pp.~21--80, Dordrecht: Springer
  Netherlands, 2017.

\bibitem{ganin_resilience_2019}
A.~A. Ganin, A.~C. Mersky, A.~S. Jin, M.~Kitsak, J.~M. Keisler, and I.~Linkov,
  ``Resilience in {Intelligent} {Transportation} {Systems} ({ITS}),'' {\em
  Transportation Research Part C: Emerging Technologies}, vol.~100,
  pp.~318--329, 2019.

\bibitem{gisladottir_resilience_2017}
V.~Gisladottir, A.~A. Ganin, J.~M. Keisler, J.~Kepner, and I.~Linkov,
  ``Resilience of {Cyber} {Systems} with {Over}- and {Underregulation},'' {\em
  Risk Analysis}, vol.~37, no.~9, pp.~1644--1651, 2017.

\bibitem{massaro_resilience_2018}
E.~Massaro, A.~A. Ganin, N.~Perra, I.~Linkov, and A.~Vespignani, ``Resilience
  management during large-scale epidemic outbreaks,'' {\em Scientific Reports},
  vol.~8, no.~1, 2018.

\bibitem{percolation}
F.~Bagnoli, E.~Bellini, E.~Massaro, and R.~Rechtman, ``Percolation and
  {Internet} {Science},'' {\em Future Internet}, vol.~11, no.~2, p.~35, 2019.

\bibitem{mcom}
S.-M. Cheng, P.-Y. Chen, C.-C. Lin, and H.-C. Hsiao, ``Traffic-{Aware}
  {Patching} for {Cyber} {Security} in {Mobile} {IoT},'' {\em IEEE
  Communications Magazine}, vol.~55, no.~7, pp.~29--35, 2017.

\bibitem{wimeh}
S.~Tanachaiwiwat and A.~Helmy, ``Encounter-based worms: analysis and defense,''
  in {\em 2006 2nd {IEEE} {Workshop} on {Wireless} {Mesh} {Networks}}, (Reston,
  VA, USA), pp.~170--172, IEEE, 2006.

\bibitem{wang}
P.~Wang, M.~C. Gonzalez, C.~A. Hidalgo, and A.-L. Barabasi, ``Understanding the
  {Spreading} {Patterns} of {Mobile} {Phone} {Viruses},'' {\em Science},
  vol.~324, no.~5930, pp.~1071--1076, 2009.

\bibitem{sheffi}
Y.~Sheffi, {\em The resilient enterprise: overcoming vulnerability for
  competitive advantage}.
\newblock Cambridge, MA: MIT Press, 2007.

\bibitem{risknet}
E.~Massaro and F.~Bagnoli, ``Epidemic spreading and risk perception in
  multiplex networks: A self-organized percolation method,'' {\em Phys. Rev.
  E}, vol.~90, p.~052817, 2014.

\bibitem{yin}
L.~Yin, Y.~Sun, Z.~Wang, Y.~Guo, F.~Li, and B.~Fang, ``Security {Measurement}
  for {Unknown} {Threats} {Based} on {Attack} {Preferences},'' {\em Security
  and Communication Networks}, vol.~2018, pp.~1--13, 2018.

\bibitem{survey}
M.~Pendleton, R.~Garcia-Lebron, J.-H. Cho, and S.~Xu, ``A {Survey} on {Systems}
  {Security} {Metrics},'' {\em ACM Computing Surveys}, vol.~49, no.~4,
  pp.~1--35, 2016.

\bibitem{ganin_multicriteria_2017}
A.~A. Ganin, D.~Marchese, Z.~A. Collier, P.~Quach, M.~Panwar, and I.~Linkov,
  ``Multicriteria {Decision} {Framework} for {Cybersecurity} {Risk}
  {Assessment} and {Management},'' {\em Risk Analysis (available online)},
  2017.

\bibitem{riskperception}
F.~Bagnoli, P.~Liò, and L.~Sguanci, ``Risk perception in epidemic modeling,''
  {\em Physical Review E}, vol.~76, no.~6, p.~061904, 2007.

\bibitem{complex}
F.~Bagnoli, E.~Bellini, and E.~Massaro, ``A {Self}-organized {Method} for
  {Computing} the {Epidemic} {Threshold} in {Computer} {Networks},'' in {\em
  Internet {Science}} (S.~S. Bodrunova, ed.), vol.~11193, pp.~119--130, Cham,
  Switzerland: Springer International Publishing, 2018.

\bibitem{compeng}
F.~Bagnoli, E.~Bellini, and E.~Massaro, ``Risk {Perception} and {Epidemics} in
  {Complex} {Computer} {Networks},'' in {\em 2018 {IEEE} {Workshop} on
  {Complexity} in {Engineering} ({COMPENG})}, pp.~1--5, IEEE, 2018.

\bibitem{virtual_inf}
J.~Ginsberg, M.~H. Mohebbi, R.~S. Patel, L.~Brammer, M.~S. Smolinski, and
  L.~Brilliant, ``Detecting influenza epidemics using search engine query
  data,'' {\em Nature}, vol.~457, no.~7232, pp.~1012--1014, 2009.

\bibitem{virtual_inf1}
D.~Scanfeld, V.~Scanfeld, and E.~L. Larson, ``Dissemination of health
  information through social networks: {Twitter} and antibiotics,'' {\em
  American Journal of Infection Control}, vol.~38, no.~3, pp.~182--188, 2010.

\bibitem{virtual_inf2}
C.~Chew and G.~Eysenbach, ``Pandemics in the {Age} of {Twitter}: {Content}
  {Analysis} of {Tweets} during the 2009 {H}1n1 {Outbreak},'' {\em PLoS ONE},
  vol.~5, no.~11, p.~e14118, 2010.

\bibitem{bagnoli_rech}
F.~Bagnoli, P.~Palmerini, and R.~Rechtman, ``Algorithmic mapping from
  criticality to self-organized criticality,'' {\em Physical Review E},
  vol.~55, no.~4, pp.~3970--3976, 1997.

\bibitem{iobt}
J.~Farooq and Q.~Zhu, ``On the secure and reconfigurable multi-layer network
  design for critical information dissemination in the internet of battlefield
  things,'' {\em IEEE Transactions on Wireless Communications}, 2018.

\end{thebibliography}
\bibliographystyle{ieeetr}

\end{document}